\documentclass[3p]{elsarticle}

\makeatletter
\def\ps@pprintTitle{%
	\let\@oddhead\@empty
	\let\@evenhead\@empty
	\let\@oddfoot\@empty
	\let\@evenfoot\@oddfoot
}
\makeatother

\usepackage{amssymb,amsmath,mathtools,bm}

\usepackage{graphicx}
\usepackage{cancel}
\usepackage{hyperref}
\usepackage[ruled,vlined]{algorithm2e}

\usepackage[colorinlistoftodos,textwidth=4cm,shadow]{todonotes}

\newcounter{Igor}

\date{}
\title {\textbf{Simulation-Guided Optimization of Granular Phononic Crystal Structure Using the Discrete Element Method}}

\author{Igor Ostanin \footnote{Corresponding author, e-mail:i.ostanin@utwente.nl}
}

\address{Multi-Scale Mechanics (MSM), Faculty of Engineering Technology, MESA+, University of Twente, P.O. Box 217, 7500 AE Enschede, The Netherlands.}

\author{Hongyang Cheng}

\author{Vanessa Magnanimo}

\address{Construction Management and Engineering- (CME), Faculty of Engineering Technology, MESA+, University of Twente, P.O. Box 217, 7500 AE Enschede, The Netherlands.}

\begin{document}

\begin{abstract}
The paper describes a novel methodology of designing granular phononic crystals for acoustic wave manipulations. A discrete element method is utilized to model the dynamics of a pulse wave propagating through the densely packed assembly of elastic spherical particles with an embedded phononic crystal - the region consisting of a certain arrangement of particles with varying densities. We suggest an optimization strategy that extremizes the useful properties of a granular phononic crystal, which are described in terms of a noise-proof functional based on frequency-wavenumber summation of spectral energy density. Few types of efficient phononic crystals are identified. The suggested methodology is of interest for a number of applications, in particular, for seismic shielding and selective sound absorption. 
\end{abstract}

\maketitle

\section{Introduction}

Understanding wave propagation in a granular medium is of great interest for multiple practical applications, including non-destructive testing \cite{Soilswaves, Clayton2011, Alva2012}, wave and shock absorption \cite{Doney2012}, noise attenuation\cite{Duvigneau2018}, spectral filtering of acoustic waves \cite{Decorated2019}, seismic shielding \cite{SeisMet2020} \textit{etc.} It is well known that the medium consisting of multiple elastic spherical particles features the effective behavior of an elastic continuum in the limit of long wavelengths and static deformations \cite{WhyEMTfails1999}, while exhibiting complex dispersive behavior in the high-frequency range, when the wavelength is comparable with sizes of constituent particles \cite{Disordered2017, Cheng2020}. In this case, each individual particle or a group of particles, confined by its neighbors, works as the resonator that can be excited by the high-frequency harmonics of the propagating acoustic perturbation and absorb part of their energy. Moreover, packings of multiple granular particles are systems with complex spectrum of eigenfrequencies what can be tuned to provide desired acoustic properties.

Granular phononic crystals and disordered structures received an increasing attention as a mean to manipulate acoustic waves. The direct problem -- analysis of the transient and steady-state regimes of wave propagation in a given granular assembly -- have been studied experimentally and numerically, including analysis of granular chains \cite{Nesterenko1984}, disordered granular medium \cite{Cheng2020}, granular phononic crystals \cite{Conical2016,Decimation2017, Decorated2019, Hexagonal2019}, soil-like multiphase systems \cite{HongyangMulti2019}, Helmholtz resonators \cite{BubbleMeta2016} \textit{etc.} However, the inverse problem -- designing the effective acoustic properties of a granular assembly by tuning particle parameters -- have gained much less attention, in spite of its obvious interest. Inverse problems have been studied in the granular matter community to achieve enhanced mechanical (strength/dilatancy) performances 
thanks to optimal particle shape, as found by neural network 
algorithms \cite{Hern2013,Jaeger_2014,Jaeger2015, Reis2015, Roth2016, Murphy2019}.  Experimental works have shown the effect of singularities, defects, 
switches on the acoustic of granular chains \cite{Boechler2011,Herbold2009,Li2014} and the potential for
vibration mitigation \cite{Gantzounis2013} while pioneering work on particle fabrication have
achieved soft acoustic metamaterials with negative index \cite{Brunet2015}. More recently, Wu et al \cite{Wu2019} have shown that mass contrast 
and arrangement can be tuned to optimise the frequency gap in a bidisperce 2D granular crystal. One-dimensional structures of hollow cylinders with tunable acoustic properties have been reported in \cite{Zhang2021}. 

In this work we suggest a systematic approach towards the designing granular structures with the desired properties. It relies on i) the discrete element method (DEM) modeling of granular assemblies of particles, ii) cost functionals based on the frequency-wavenumber summation of spectral energy density, and iii) numerical simulation-guided optimization procedure. DEM \cite{Cundall_Geo_1979} is the standard numerical tool to model static and dynamic properties of granular assemblies. In this work we combine its strong features with the technique of unbounded gradient-free optimization \cite{2020SciPy}, in order to obtain desirable dispersive properties of a phononic crystal. It worth noting here that the optimization approaches are often used in DEM for the purposes of microscopic calibration of contact models \cite{grainLearning,Genetic_2018}, however, so far they have not yet been applied for designing taylored macroscopic acoustic properties of a granular assembly.

Our paper is organized as follows. The Section 2 gives details on the numerical techniques and methods used in our work. Section 3 discusses the phononic crystals that have been obtained using our techniques, as well as their acoustic properties. Section 4 places our results into a general context of the research efforts in the field and compares our phononic structures with the known ones. Section 5 summarizes the contributions of our work.          

\section{Method}

\subsection{Discrete element method} We utilize DEM to study the phenomenon of wave propagation in a granular solid. YADE open source DEM package \cite{Yade} is used in our computations. The method computes the dynamics of rigid spherical bodies. In this work we use equal-sized spherical particles with mass $M$, radius $R$ and moment of inertia $\frac{2}{5} M R^2$, using velocity Verlet time integration scheme with the fixed time step. Bodies interact via Hertz-Mindlin no-slip contact model, providing force and moment responses in a limit of vanishingly small overlaps between particles. For two spheres in contact with the normal overlap $\mathbf{u}_n$, an incremental tangential displacement $\Delta \mathbf{u}_s$, relative rotational angle $\boldsymbol{\theta_c}$ at the contact point, the interparticle normal force $\mathbf{F}_n$, incremental tangential force $\Delta \mathbf{F}_s$ and contact moment $\mathbf{M}_c$ are given by:

\begin{equation} \label{1}
	\mathbf{F}_{n}=\frac{2E_{c}\mathbf{u}_{\mathit{n}}}{3(1-\nu_{c}^{2})}\sqrt{\frac{R\left|\mathbf{u}_{n}\right|}{2}}
\end{equation}

\begin{equation} \label{2}
	\Delta \mathbf{F}_s = \frac{2 E_c \Delta \mathbf{u}_s}{(1+\nu_c)(2-\nu_c)}\sqrt{\frac{R\left|\mathbf{u}_{n}\right|}{2}}
\end{equation}

\begin{equation} \label{3}
	\left| \mathbf{F}_s \right| \le \tan{\mu_f}  \left| \mathbf{F}_n \right|
\end{equation}

\begin{equation} \label{4}
	\mathbf{M}_c = k_m \boldsymbol{\theta_c}
\end{equation}

\begin{equation} \label{5}
	\left|\mathbf{M}_c\right| \le R \eta_m \left| \mathbf{F}_n \right| 
\end{equation}

where $E_c$ and $\nu_c$ are the contact model Young's modulus and Poisson's ratio, $\mu_f$ is the interparticle friction angle, $k_m$ is the rolling stiffness, $\eta_m$ is the rolling friction coefficient, that controls the plastic limit of $\mathbf{M}_c$. The normal force $\mathbf{F}_n$ is a function of and overlap $\mathbf{u}_n$, whereas the shear force $\mathbf{F}_s$ is accumulated at every time step and is checked against the limit \ref{3}.  

\begin{figure}
	\includegraphics[width=17cm]{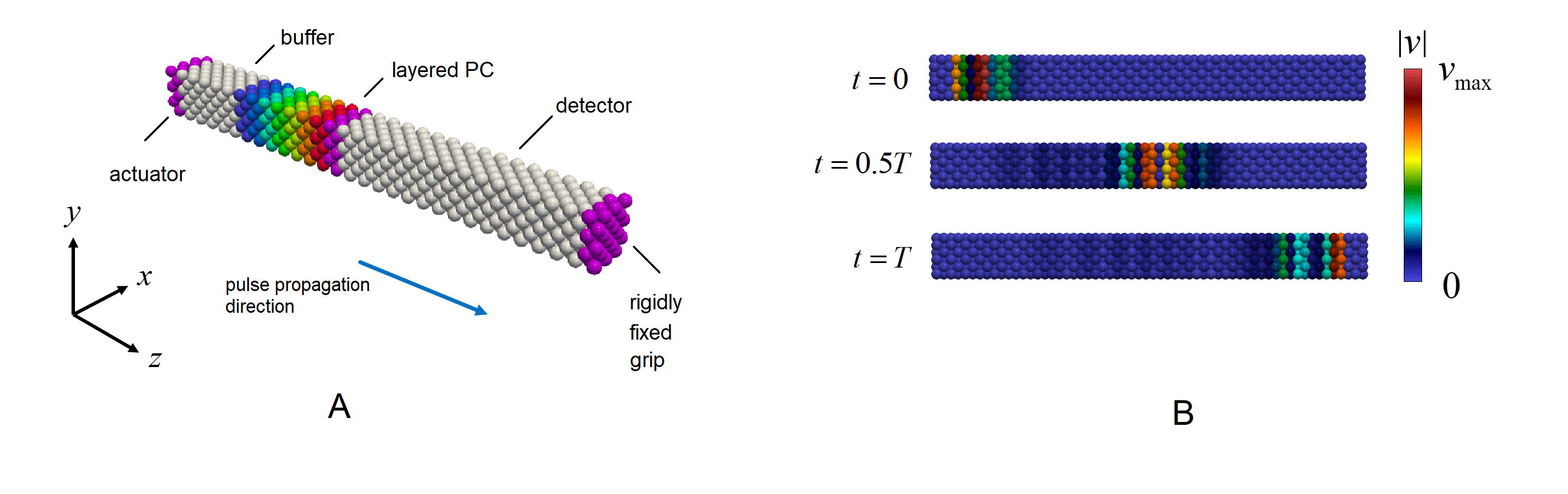}
	\protect\caption{(A) Numerical experiment schematics (B) Propagation of a P-wave pulse through a homogeneous granular column.}
\end{figure}

\subsection{Numerical experiment setup} The simulation domain is the long column of close-packed particles, incorporating $K \times K \times N$ face-centered cubic (FCC) cells, with $N \gg K$ (Fig.1(A)). Width, height, and length of the column are aligned with $x$, $y$ and $z$ Cartesian axes, correspondingly. The column is initially placed into periodic boundary conditions along every axis, and, using standard YADE triaxial controller \cite{Yade}, is gradually compressed until reaching a prescribed isotropic state of stress. The excess of kinetic energy is taken out by local damping, imposing forces opposite to velocities and proportional to forces acting on particles, as described in \cite{Yade}. For regular grids and equal-sized particles of same stiffness, it is possible to solve explicitly for an equilibrium strain associated with a given state of stress. However, the numerical equilibration step is preserved to keep the approach general. A local damping proportionality coefficient of $0.2$ is used during specimen relaxation.

The domain is then decomposed into layers along $z$, of size $K \times K \times 1$ cells each. Once the arrangement of particles is equilibrated under compression, periodic boundary conditions along $z$ are replaced with an unbounded domain, while $x$ and $y$ periodic conditions are preserved. Simultaneously, two layers at the edges of the column are rigidly fixed so that the state of stress within the column is not perturbed. 

The interior $N-2$ layers of the column are divided onto $N_b$ buffer layers, $N_c$ phononic crystal (hereafter referred to as PC) layers and $N_d$ detector layers. The density of the buffer layers and detector layers are constant, while the densities of particles within PC are varied to optimize its useful properties - every layer of PC region consists of particles of the same density. 

At the beginning of the test, the local damping is set to zero, and a short pulse of P- or S-wave is agitated at the left edge of the column, by prescribing the $z$- or $y$-component of displacement of the left rigidly fixed layer (actuator). We agitate a single sine period $u_{z}(t) = A \sin{\omega_{p}t}$ for P-wave or $u_{y}(t) = A \sin{\omega_{p}t}$ for S-wave, with the magnitude of particle displacement $A$ much smaller than the equilibrium overlaps, leading to linear regime of oscillations. The frequency of initial sine pulse is chosen $\omega_{p} = 0.2\omega_{0}$, where $\omega_{0}$ is the highest eigenfrequency of the system for a given confining stress. Such a pulse creates a rich spectrum of a frequencies in the signal. The pulse travels along the buffer region of the column, exhibiting dispersion inherent to discrete spring-mass system \cite{Ruzzene2014}. After reaching the PC region, the pulse partly reflects from the crystal, partly absorbs by it, exciting multiple eigenmodes, and partly propagates through, reaching the detector region.

\begin{figure}
	\includegraphics[width=17cm]{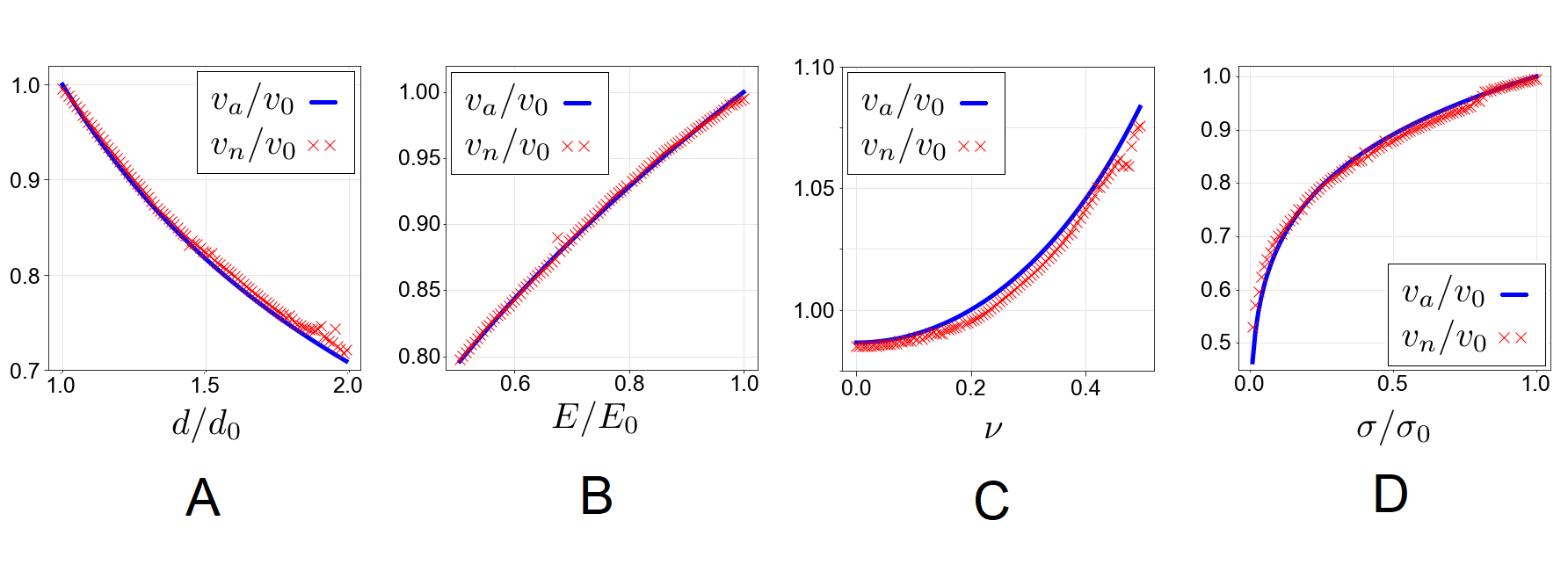}
	\protect\caption{Theoretical (blue line) and numerically computed (red marks) dependencies of P-wave velocity in a homogeneous granular column as functions of density (A), Young's modulus (B) and  Poisson's ratio (C) of spherical particles, as well as the confining stress (D), in case of zero friction.}
\end{figure}

As a validation step, we study P-wave pulse propagation in a granular column in the case of equal densities of the PC layers and zero friction (Fig. 1(B)). Such a case is well studied and admits simple analytical solution through linearization of Hertzian contact forces for a given confining stress \cite{Goddard1990,Coste1997}. The scaling of wave velocity with respect to density, Young's modulus, Poisson's ratio and confining stress is given by:

\begin{equation} \label{6}
	v_a(d, E, \nu, \sigma) = C  {\sigma}^{1/6} {d}^{-1/2} \left( {\frac{E}{2(1-\nu^2)}}\right) ^{1/3}
\end{equation}

The constant $C$ depends on the frequency of the propagating signal, strain in the system and packing type. Fig. 2 gives the wave velocity \ref{6}, as compared to the group wave propagation velocity $v_n(d, E, \nu, \sigma)$, measured in the simulation. The latter one is determined using the linear approximation of peak position vs time dependency. Both velocities are studied as functions of particle's material density (Fig.2(A)), Young's modulus (Fig. 2(B)), Poisson's ratio (Fig.2(C)) and the confining stress (Fig.2(D)). It appears that all the scalings follow closely the theoretical prediction \ref{6}, which confirms the validity of our model.

\subsection{Dispersion measurements} In order to characterize the spectral properties of the pulse and their modification by the granular PCs, we use temporal-spatial Fourier transform of the velocity field, which gives the description of the pulse in terms of spectral energy densities in frequency-wavenumber domain. In this work we study only regular arrangements of particles, which allows us to use fast Fourier transform (FFT) both in space and time. 
The arrangement of $K \times K \times N$ face-centered cubic (FCC) cells is separated into $2N$ layers of equal thickness along the length of the column in such a way that the center of each layer corresponds to longitudinal coordinates of the $N_k=2K^2$ particles. The normal and transversal velocity components of layer $k$ are obtained by averaging over particles belonging to the layer:

\begin{equation} \label{7}
	\begin{split}
		v_k(t)^{\parallel} &= \frac{1}{N_k} \sum_{i}^{N_k} v_i(t)^{\parallel}, \\
		v_k(t)^{\perp} &= \frac{1}{N_k} \sum_{i}^{N_k} v_i(t)^{\perp}, \\
	\end{split}
\end{equation}

here $v_i(t)^{\parallel}$ and $v_i(t)^{\perp}$ are $z$- and $y$- components of the velocity of the particle (for symmetry reasons, $x$- component of the velocity is zero). 

By discretizing the temporal evolution of every layer's velocity into $2M$ equal timesteps as $v_{kl}^{\parallel} = v_k(l\Delta t)^{\parallel}$,$v_{kl}^{\perp} = v_k(l\Delta t)^{\perp}$, $k = 1..2N$, $l = 1..2M$, we obtain $2N \times 2M$ 2D space-time velocity array. The temporal $(t_{min}, t_{max})$ and spatial $(z_{min}, z_{max})$ window in this array is chosen such that it covers either $N_d$ detector layers at the time span when the pulse peak travels though (``detector'' FFT), or all the layers during the complete simulation time (``full'' FFT). The chosen space-time window is then transformed into $N \times M$ complex-valued frequency-wavenumber array according to standard 2D FFT procedure for a real-valued signal [\cite{numpy, numpy_fft}]:

\begin{equation} \label{7}
	\begin{split}
		\tilde{v}_{ij}^{\parallel} &=\tilde{v}(i\Delta \omega, j \Delta \kappa)^{\parallel} = \sum_{k} \sum_{l} v_{kl}^{\parallel} e^{ - \pi I (ik/N+jl/M)}, \\
		\tilde{v}_{ij}^{\perp} &=\tilde{v}(i\Delta \omega, j \Delta \kappa)^{\perp} = \sum_{k} \sum_{l} v_{kl}^{\perp} e^{ - \pi I (ik/N+jl/M)}, \\  
	\end{split}
\end{equation}

Here $\Delta \omega = 2\pi/(t_{max}-t_{min})$, $\Delta \kappa = 2\pi/(z_{max}-z_{min})$.

We use spectral energy densities $S^l(\omega, \kappa) = \left| \mathbf{\tilde{v}}(\omega, \kappa)^{\parallel} \right|^2$ and $ S^t(\omega, \kappa) = \left| \mathbf{\tilde{v}}(\omega, \kappa)^{\perp} \right|^2$ to characterize the dispersive properties of our granular structures for P- and S-waves, respectively. Fig. 3 presents the ``full'' FFT spectral energy distributions $S^l(\omega, \kappa)$ for a P-wave propagating in an FCC column consisting of $3 \times 3 \times 30$ (Fig. 3(A)) and $3 \times 3 \times 120$ (Fig. 3(B)) FCC cells of frictionless Hertzian beads. The dispersion branches are clearly identifiable. The linearized theory of dispersion for this case \cite{Ruzzene2014} predicts the dispersion branch of the shape: 

\begin{equation} \label{10}
	\Omega = \sqrt{2(1-\cos{\mu})}
\end{equation}

where normalized frequency $\Omega \in (0,2)$ and wavenumber $\mu \in (0,\pi)$ define the first Brillouin zone of the dispersion relation. Below we will utilize this normalization for spectrograms $S^l(\Omega, \mu),S^t(\Omega, \mu)$.

\begin{figure}
	\includegraphics[width=17cm]{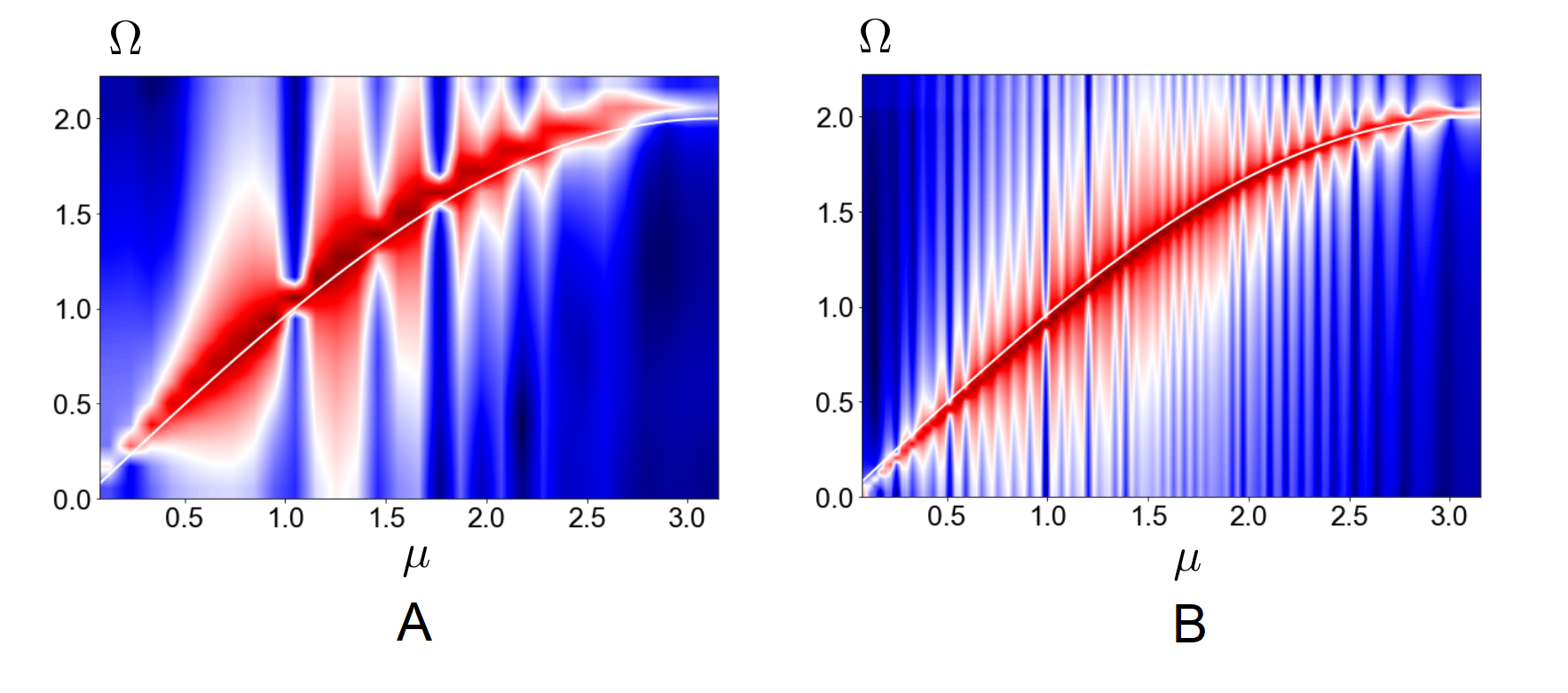}
	\protect\caption{Dispersion branches in short(A) and long(B) columns. Here and below we use seismic (``cold'' to ``warm'') colormaps and logarithmic color legends for spectral energy densities to make fine features of spectrograms clearly identifiable.}
\end{figure}

\subsection{Optimization technique} The primary goal of our work is to substantially modify the acoustic properties of the granular column by varying the properties of its constituent particles. In order to extremize the acoustic properties of a PC without perturbing static force equilibrium, we only vary densities of PC layers, while leaving their stiffnesses unchanged. The densities $d_i$ of PC layers are varied independently in an unbounded and unconstrained multidimensional optimization procedure. The optimizer seeks for a vector of unknowns $\mathbf{X}: X_i \in \mathcal{R}$, which are mapped to $ \mathbf{d}: d_i \in [d_{min}, d_{max}]$ in the following way:

\begin{equation} \label{12}
	\mathbf{d}(\mathbf{X}) = \frac{d_{max}+d_{min}}{2} - \frac{d_{max}-d_{min}}{2} \cos{\mathbf{X}}
\end{equation}

The densities $[d_{min}, d_{max}]$ are related with the density of surrounding regions $d_{0}=(d_{min} + d_{max})/2$. A simulation-guided optimization procedure is employed to calibrate the densities of particles of a PC in order to extremize the cost functional. Simulation results inevitably involve noise, therefore we deal with the noisy optimization problem. In order to decrease the influence of noise, it is preferable to use a certain filtering/averaging technique, in order to construct noise-proof, locally smooth functionals. 

Assume we would like to minimize the spectral density in a given domain $\Gamma$ in $\Omega-\mu$ space. In order to achieve that, we can minimize the following functionals:

\begin{equation}\label{16}
	\begin{split}
		\mathcal{L}^{l}(\mathbf{X}, \Gamma) &= \iint_\Gamma S^{l}(\Omega, \mu) \,d \mu\,d\Omega , \\
		\mathcal{L}^{t}(\mathbf{X}, \Gamma) &= \iint_\Gamma S^{t}(\Omega, \mu) \,d \mu\,d\Omega.
	\end{split}
\end{equation} 

Below we'll utilize the particular kind of such functionals with domain $\Gamma$ being the rectangle: $\Gamma(\Omega, \mu): \mu \in (\mu_{min},\mu_{max}), \Omega \in (\Omega_{min},\Omega_{max})$. The functional is therefore defined as:

\begin{equation} \label{17}
	\begin{split}
		\mathcal{L}^{l}(\mathbf{X}, \Gamma) &= \int_{\mu_{min}}^{\mu_{max}} \int_{\Omega_{min}}^{\Omega_{max}} S^{l}(\Omega, \mu) \,d\mu\,d\Omega, \\
		\mathcal{L}^{t}(\mathbf{X}, \Gamma) &= \int_{\mu_{min}}^{\mu_{max}} \int_{\Omega_{min}}^{\Omega_{max}} S^{t}(\Omega, \mu) \,d\mu\,d\Omega.
	\end{split}
\end{equation}

We define the normalized version of this functional as:

\begin{equation} \label{18}
	\begin{split}
		\mathcal{\hat{L}}^{l}(\mathbf{X}, \Gamma) &= \mathcal{L}^{l}(\mathbf{X},\Gamma)/\mathcal{L}^{l}(\mathbf{X}_0,\Gamma), \\
		\mathcal{\hat{L}}^{t}(\mathbf{X}, \Gamma) &= \mathcal{L}^{t}(\mathbf{X},\Gamma)/\mathcal{L}^{t}(\mathbf{X}_0,\Gamma), \\
	\end{split}
\end{equation}

In the examples below, $\mathbf{X}_0$ and the vector of initial densities $\mathbf{d}(\mathbf{X}_0)$ are given as:    

\begin{equation} \label{15}
	\begin{split}
		\mathbf{X}_0 &= \frac{\pi}{2} \mathbf{I}, \\
		\mathbf{d}(\mathbf{X}_0) &= \frac{d_{max}+d_{min}}{2} \mathbf{I} 
	\end{split}
\end{equation}

We use the measure $\delta$ quantifying the quality of the results of the optimization procedure:

\begin{equation} \label{19}
	\delta = \mathcal{\hat{L}}(\mathbf{X}_f), 
\end{equation}

where $\mathbf{X}_f$ is the final set of optimization parameters at the end of optimization procedure.

In order to perform an optimization of the PC parameters in multidimensional space, we have tested few different derivative-free optimization techniques, immediately available as parts of SciPy Optimize library \cite{2020SciPy}, part of Anaconda Python distribution - Powell, Nelder-Mead, COBYLA \textit{etc.} Powell's conjugate direction method \cite{Powell1964} has been chosen as the most efficient for our problem formulation - it displayed reasonably fast convergence and low sensitivity to noise, often leaving other algorithms trapped in local minima. This method allows to find a local minimum of a continuous function without computing derivatives. A more advanced stochastic optimization techniques have been developed recently in the context of the calibration of DEM models \cite{grainLearning} however, in our case the explicit optimization techniques appear to be more suitable, since the noise level still ensures smoothness of the cost functional with respect to search parameters, which allows using a relatively simple toolkit of local minimization.

\begin{table}
	\label{tab1}
	\caption{List of the optimization runs and their parameters. The columns give the test number, number of layers $N$, number of buffer layers $N_b$, number of PC region layers $N_c$, number of detector region layers $N_d$, type of functional, wave mode ("P" - pressure wave, "S" - shear wave), and the normalized final value of the cost functional $\delta$ for the case of layered structures.}
	\medskip
	\centering
	\begin{tabular}{|c|c|c|c|c|c|c|c|}
		\hline 
		Run & $N$ & $N_b$ & $N_c$ & $N_d$ & func. & w.m. & $\delta$ (layers)   \tabularnewline
		\hline
		\hline 
		1 & 50 & 2 & 8 & 38 & $\mathcal{\hat{L}}^{l}(\mathbf{X}, \Gamma_1)$ & P & $0.592$   \tabularnewline
		\hline
		2 & 50 & 2 & 8 & 38 & $\mathcal{\hat{L}}^{l}(\mathbf{X}, \Gamma_2)$ & P & $0.530$  \tabularnewline
		\hline 
		3 & 50 & 2 & 8 & 38 & $\mathcal{\hat{L}}^{l}(\mathbf{X}, \Gamma_3)$ & P & $0.494$  \tabularnewline
		\hline 
		4 & 50 & 2 & 8 & 38 & $\mathcal{\hat{L}}^{l}(\mathbf{X}, \Gamma_4)$ & P & $0.039$  \tabularnewline
		\hline 
		5 & 50 & 2 & 8 & 38 & $\mathcal{\hat{L}}^{l}(\mathbf{X}, \Gamma_5)$ & P & $0.221$  \tabularnewline
		\hline
		\hline	 
		6 & 50 & 2 & 16 & 30 & $\mathcal{\hat{L}}^{l}(\mathbf{X}, \Gamma_1)$ & P & $0.478$  \tabularnewline
		\hline 
		7 & 50 & 2 & 16 & 30 & $\mathcal{\hat{L}}^{l}(\mathbf{X}, \Gamma_2)$ & P & $0.399$  \tabularnewline
		\hline 
		8 & 50 & 2 & 16 & 30 & $\mathcal{\hat{L}}^{l}(\mathbf{X}, \Gamma_3)$ & P & $0.184$  \tabularnewline
		\hline 
		9 & 50 & 2 & 16 & 30 & $\mathcal{\hat{L}}^{l}(\mathbf{X}, \Gamma_4)$ & P & $0.102$  \tabularnewline
		\hline 
		10 & 50 & 2 & 16 & 30 & $\mathcal{\hat{L}}^{l}(\mathbf{X}, \Gamma_5)$ & P & $0.408$  \tabularnewline
		\hline
		\hline		
		11 & 50 & 2 & 8 & 38 & $\mathcal{\hat{L}}^{t}(\mathbf{X}, \Gamma_1)$ & S & $0.506$  \tabularnewline
		\hline 
		12 & 50 & 2 & 8 & 38 & $\mathcal{\hat{L}}^{t}(\mathbf{X}, \Gamma_2)$ & S & $0.700$  \tabularnewline
		\hline 
		13 & 50 & 2 & 8 & 38 & $\mathcal{\hat{L}}^{t}(\mathbf{X}, \Gamma_3)$ & S & $0.798$  \tabularnewline
		\hline 
		14 & 50 & 2 & 8 & 38 & $\mathcal{\hat{L}}^{t}(\mathbf{X}, \Gamma_4)$ & S & $0.014$  \tabularnewline
		\hline 
		15 & 50 & 2 & 8 & 38 & $\mathcal{\hat{L}}^{t}(\mathbf{X}, \Gamma_5)$ & S & $0.047$  \tabularnewline
		\hline
		\hline
		16 & 50 & 2 & 16 & 30 & $\mathcal{\hat{L}}^{t}(\mathbf{X}, \Gamma_1)$ & S & $0.574$  \tabularnewline
		\hline 
		17 & 50 & 2 & 16 & 30 & $\mathcal{\hat{L}}^{t}(\mathbf{X}, \Gamma_2)$ & S & $0.289$  \tabularnewline
		\hline 
		18 & 50 & 2 & 16 & 30 & $\mathcal{\hat{L}}^{t}(\mathbf{X}, \Gamma_3)$ & S & $0.202$  \tabularnewline
		\hline 
		19 & 50 & 2 & 16 & 30 & $\mathcal{\hat{L}}^{t}(\mathbf{X}, \Gamma_4)$ & S & $0.042$  \tabularnewline
		\hline 
		20 & 50 & 2 & 16 & 30 & $\mathcal{\hat{L}}^{t}(\mathbf{X}, \Gamma_5)$ & S & $0.244$  \tabularnewline
		\hline   
	\end{tabular}
\end{table}

\section{Results}

\begin{figure}
	\includegraphics[width=17cm]{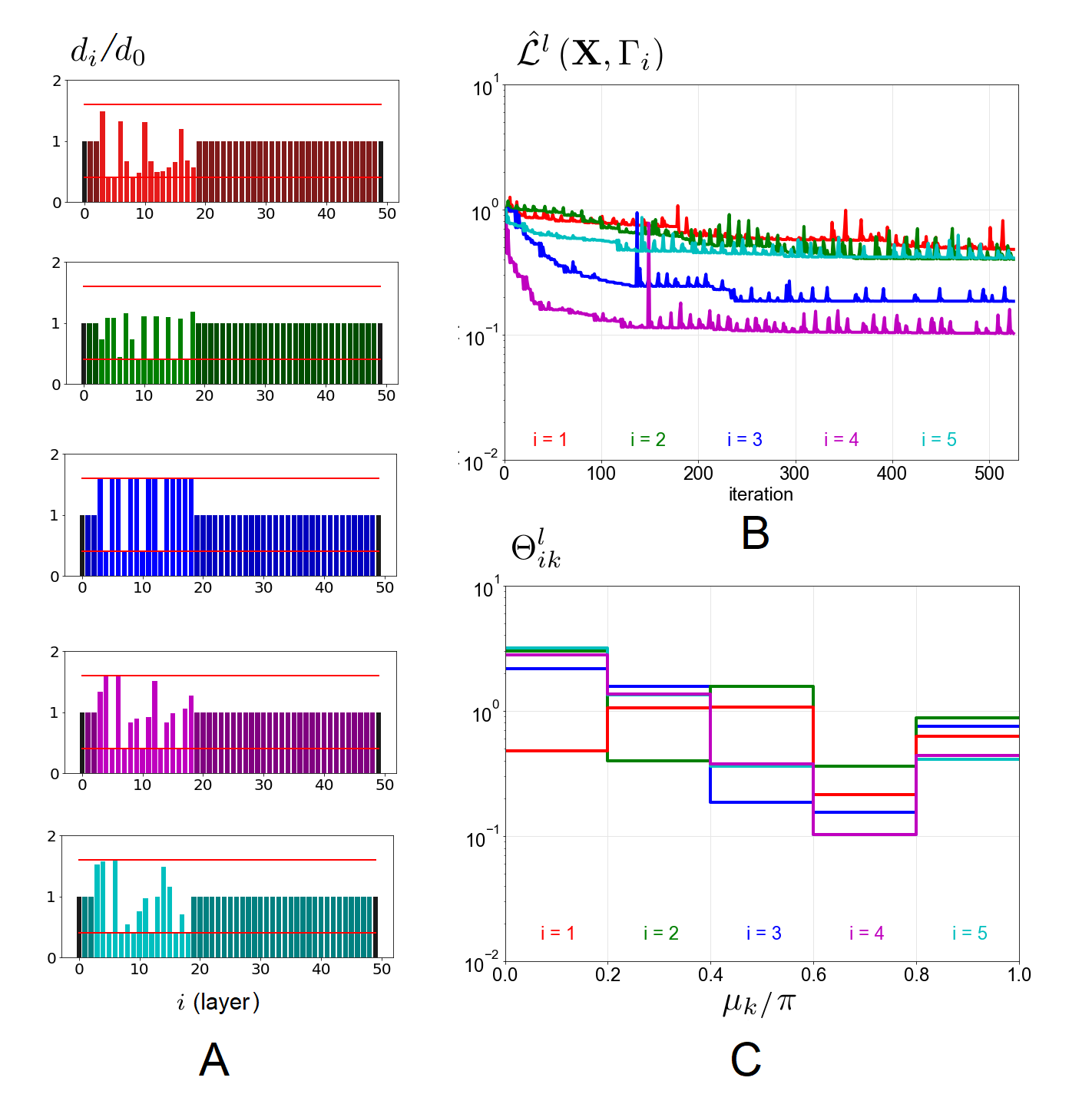}
	\protect\caption{Band energy optimization, P-wave, 16-layers PC (tests 6-10). (A) density distributions, (B) evolution of functionals, (C) averaged spectral energy density for optimized structures.}
\end{figure}

\begin{figure}
	\includegraphics[width=17cm]{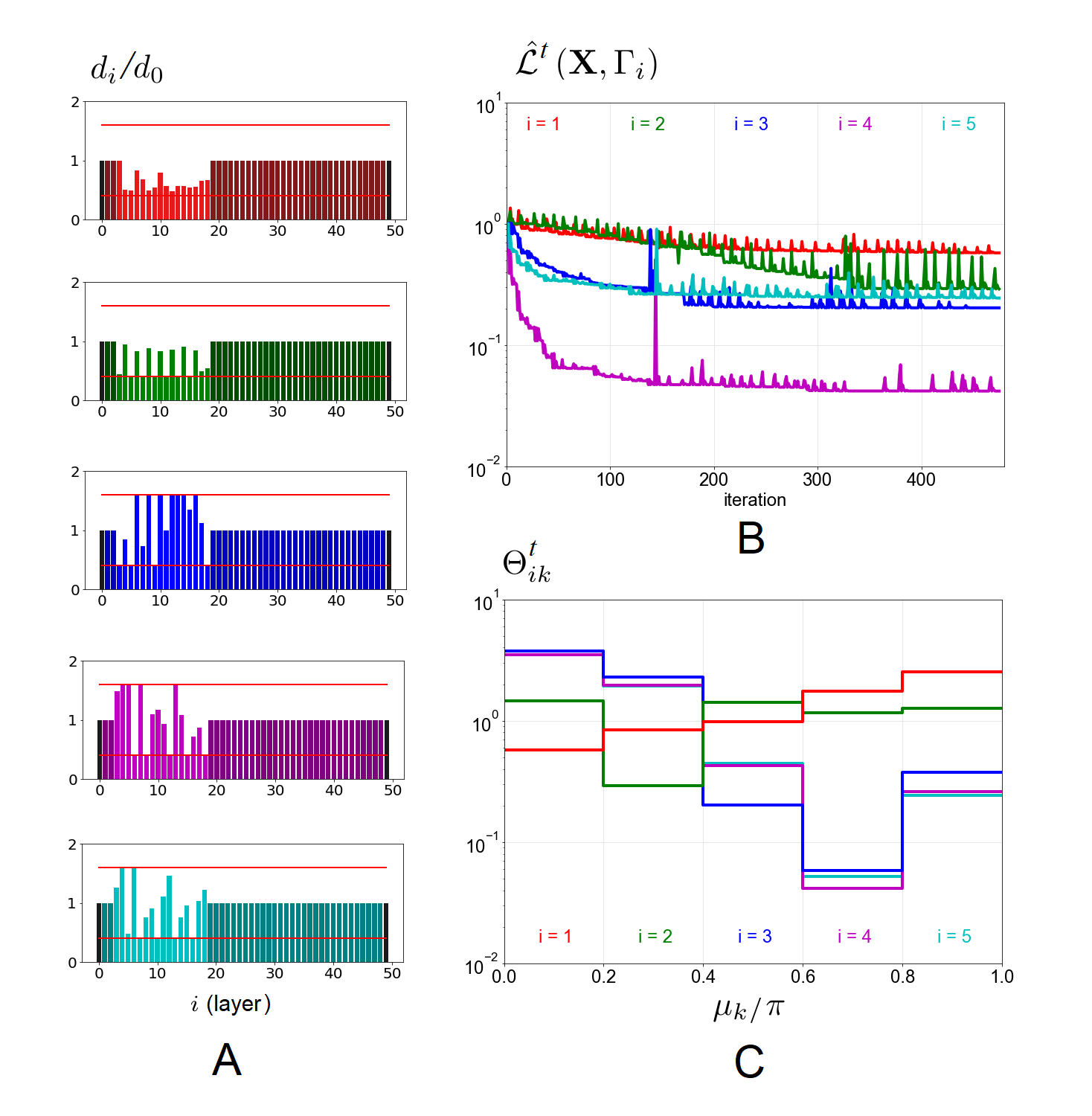}
	\protect\caption{Band energy optimization, S-wave, 16-layers PC (tests 16-20). (A) density distributions, (B) evolution of functionals, (C) averaged spectral energy density for optimized structures.}
\end{figure}

We apply the above-described technique to study the influence of density variations in layered granular structures on their spectral properties. The densities of the layers are varied to minimize the cost functionals.

As has been discussed above, P- or S-wave is initiated in a column with the layered structure: one actuator layer (first grip) is followed by $N_b$ buffer layers, then follow $N_c$ layers of a PC, $N_d$ layers of a detector region and a final terminating layer (second grip). The densities of grips, buffer and detector layers remain the same, while the densities of a PC are varied within the limits $d_{min} = 0.4 d_0$, $d_{max} = 1.6 d_0$. The results of all optimization experiments are summarized in Table 1.

In order to directly optimize layered structures with the tailored spectral properties, we use the functional in eq. \ref{18} as the measure of quality of the PC structure. The first Brillouin zone of a baseline structure, given by the rectangle $\Omega \in [0,2)$, $\mu \in [0,\pi)$ is divided into 5 equal stripes along the wavenumber direction:

\begin{equation}\label{20}
	\Gamma_i: \Omega \in [0,2), \mu \in [0.2(i-1),0.2i)\pi, i = 1..5
\end{equation} 

The results of series of optimization experiments with 16-layers phononic crystals and the cost functionals $\mathcal{\hat{L}}(\mathbf{X}, \Gamma_i)$, $i = 1..5$ are presented in Fig. 4 (P-wave) and Fig. 5 (S-wave). For Fig. 4 and Fig. 5, 
Panels (A) give the distribution of densities along the length of the column. The region of PC is indicated with the lighter color. Panels (B) give the evolution of the cost functionals in the course of optimization procedure. Each peak of the functional is associated with the bidirectional procedure of 1D minimization with respect to a single search vector in a parameter space -- a peculiarity of Powell's conjugate direction algorithm. Panels (C) give the values of spectral energy densities averaged over the domains $\Gamma_k$, normalized on the corresponding averages of a baseline structure, for every optimization band $\Gamma_i$:

\begin{equation}\label{21}
	\begin{split}
		\Theta_{ik}^{l} &= \mathcal{\hat{L}}^{l}(\mathbf{X}_f^i, \Gamma_k)/ \mathcal{\hat{L}}^{l}(\mathbf{X}_0, \Gamma_k), \\
		\Theta_{ik}^{t} &= \mathcal{\hat{L}}^{t}(\mathbf{X}_f^i, \Gamma_k)/ \mathcal{\hat{L}}^{t}(\mathbf{X}_0, \Gamma_k)
	\end{split}
\end{equation} 

Here $\mathbf{X}_f^i$ is the set of final parameters achieved in optimization with respect to wavenumber stripe $\Gamma_i$.

One can clearly see from Fig. 4,5 that the optimization with respect to minimum energy density in a chosen region of the spectrogram indeed decreases the energy density in this region and increases it in the other areas of the spectrogram.
It is very important to note that the found optimal density distributions remain nearly unchanged when averaged over last 5, 10 or 100 optimization iterations. This clearly indicates that the found solutions are robust, and gradient-free Powell algorithm does a good job finding locally optimal solutions for our optimization problem formulation.

\begin{figure}
	\includegraphics[width=17cm]{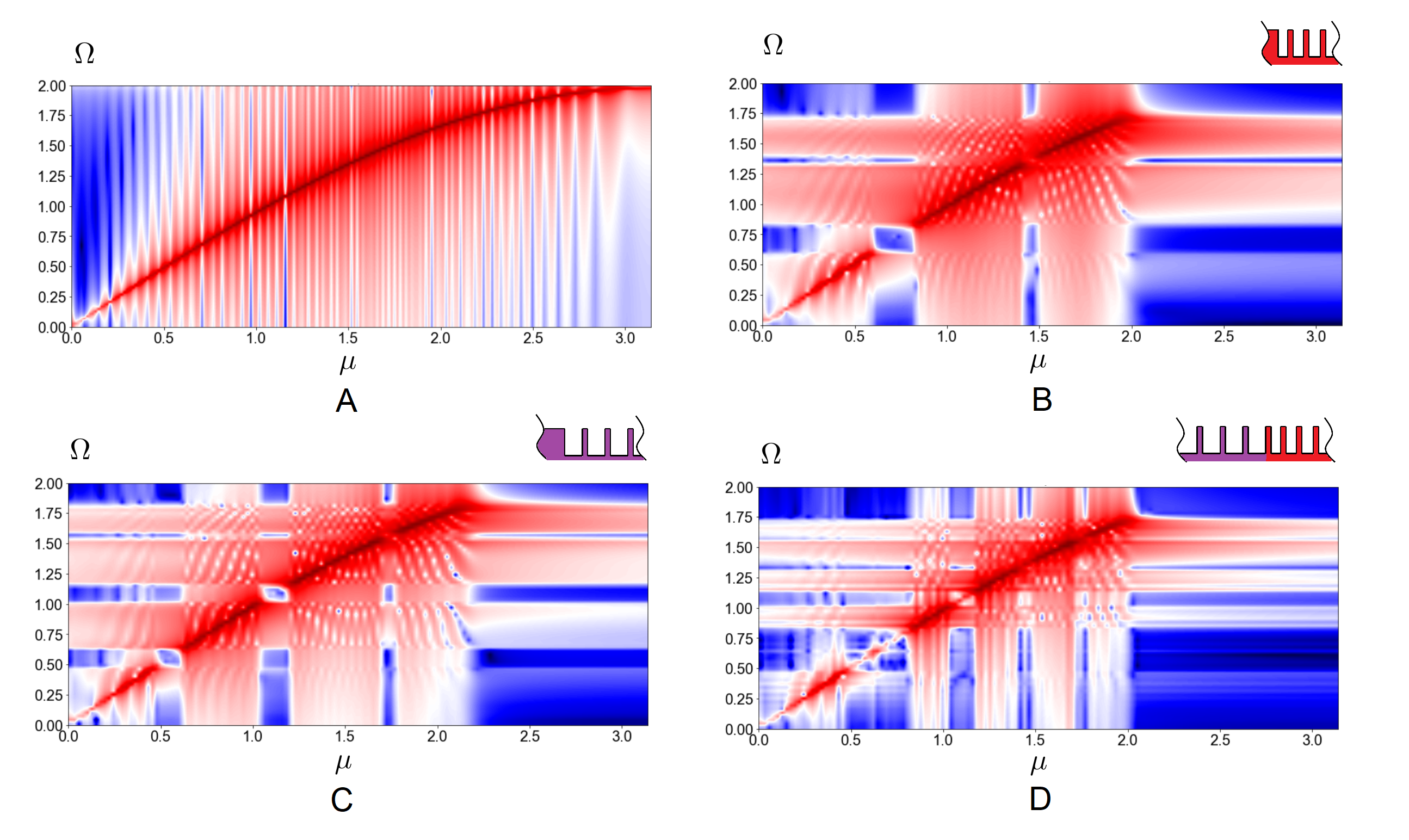}
	\protect\caption{Upscaled "meander" structures ("detector" FFTs). (A) plain background density of PC (B) Light inclusion $d=0.4$, with every 3-rd layer of $d=1$ (C) Light inclusion $d=0.4$, with every 4-th layer $d=1.0$ (D) Sequential arrangement halves of crystals (B) and (C). For (B-D), the described structures of phononic crystal layers are sketched above the spectrograms.}
\end{figure}

\subsection{Larger scale simulation -- ``meander'' structure}

Our optimization results indicated that certain patterns of density distribution often occur in the result of the optimization.  In this subsection we will consider properties of one of them, which we will call ``meander'', to demonstrate the possibilities offered by the  granular layers manipulation. This density arrangement is naturally found by the optimization algorithm as an ultimate low-frequency filter. It is a relatively light inclusion with density $d_{min} = 0.4$, complemented with periodic arrangement of layers with density close to $d_{0} = 1.0$. (See, e.g. Fig. 4,5(A), stripes 1 and 2). 

Fig. 6 summarizes the property of such structures, upscaled to the sizes four times larger than ones used in the optimization procedure -- every tested specimen had $N = 200$, $N_b = 2$, $N_c = 64$. 

Fig. 6(A) demonstrates the dispersion branch in absence of heterogeneity of PC, as obtained using FFT for $N_d$ ``detector'' layers only. The spectral energy density is evenly distributed along the branch of a dispersion relation. Fig. 6(B-D) demonstrate the effect of different ``meander'' filters on the dispersion relation. We can see that the effect of homogeneity can be split in two parts. First, all the meandered structures behave as high-frequency filters, eliminating propagation of the frequencies approaching highest eigenfrequencies of in the system. This effect is rather predictable, since fine-level resonators explicitly introduce new absorption frequencies in a higher range. The more interesting is the second effect - emergence of the suppression bands in a low-frequency range. The suppression is nearly complete - the difference between blue and red color on the spectrograms in Fig. 6 is seven to eight orders of magnitude. Fig. 6(B-D) demonstrates that manipulation of the ``meander'' parameters allows to tune the position of absorption bands, and sequential layering of these filters results in a superposition of the effects, allowing to engineer broad bands in wide frequency ranges.

\section{Discussion}
\label{disc}

As we could see from the previous section, the proposed approach is promising for designing granular PCs with tunable attenuation and spectral properties. Within the simple tested formulations and regular geometries of the PC, the simulation-guided optimization technique appears to be capable of finding non-trivial solutions that locally extremize integrals of spectral energy density over a given frequency-wavenumber window. As Fig. 4,5(C) suggest, provided optimal solutions minimize the energy in requested wavenumber bands, affecting other bands to a lesser extent.

We should note that the optimized configurations do affect the energy distribution beyond the energy band they were optimized for, especially -  close to multiples of optimized wavenumber bands - simply because of the fundamental properties of oscillating systems. The questions related to simultaneous optimization of spectral energy density in multiple regions of frequency-wavenumber domain remain outside of the scope of this paper.   

It worth noting that the results that we obtain are the local minima, which is supported by both termination criteria of the optimization procedure and negligible dependence of the functional value and configurations found on the last iterations.   The solutions are likely to be global minima as well - the optimization procedures that initiated with random guesses often converge to the same solution. However, the global exploration of parameter space remains computationally expensive even for models of moderate sizes considered in this paper. 

It is our experience that similar optimization techniques applied to random PC structures appear to be rather inefficient due to complex noisy and non-monotonous dependence of cost functionals. The more advanced methods of noisy optimization, \textit{e.g.} the ones based on sequential Monte-Carlo techniques \cite{grainLearning} may in principle be used for similar optimization of composition of random granular structures for desirable acoustic properties. However, this brings the necessity of all-particle discrete 3D Fourier transform, representatively large specimens, multiple random realizations at every optimization iteration, large numbers of iterations and expensive time integration spans.

The upscaled low-parametric structures studied have predictably demonstrated good damping/filtering properties. Our framework in principle allows tuning such structures by adjusting very few particle parameters.

Clearly, the potential of the described approach is significantly higher than simple examples showcased above. Given sufficient high-performance computing resources one can achieve optimization of much larger and detailed systems with respect to more complicated cost functionals. In particular, the technique can be used for optimization of nonlinear properties of granular arrangements within and beyond Hertzian approximation, as well as for the the case of multiphase systems, which promises a very broad range of applications in multiple applied areas including composites and metamaterials, soil mechanics, biotechnology and medicine.

\section{Conclusion}

In this work we have suggested a new approach to design and optimization of the acoustic properties of granular phononic crystals. A discrete element method simulation was used in combination with out-of-the-box multidimensional optimization algorithms to extremize the useful properties of layered and random phononic crystal structures. Clearly, suggested methodology can be generalized and developed in few possible directions. Employing high-throughput and high-performance computing resources would allow to study both much larger models and more involved optimization formulations. 
Demonstration optimization codes are available at \url{https://bitbucket.org/iostanin/pco_framework}

\section*{Acknowledgements} 

I.A.O. postdoc project was supported by the Materials programme of the University of Twente, hosted by MESA+. The financial support from the Dutch Research Council (NWO) through OpenMind Project "Soft Seismic Shields" is gratefully acknowledged.

\bibliographystyle{unsrtnat}
\bibliography{manuscript}

\end{document}